\documentclass[runningheads]{llncs}

\usepackage{graphicx}
\usepackage{booktabs}
\usepackage{algorithm}
\usepackage{algpseudocode}
\usepackage{ulem}
\usepackage{pifont}

\usepackage[T1]{fontenc}
\usepackage{graphicx}
\usepackage[colorlinks=true, allcolors=blue]{hyperref}
\begin{document}
\title{Robustly Optimized Deep Feature Decoupling Network for Fatty Liver Diseases Detection}
\titlerunning{Robust Optimization for Fatty Liver Disease Detection}
\author{Peng Huang\inst{1} 
\and Shu Hu\inst{2} 
\and Bo Peng\inst{1} 
\and Jiashu Zhang\inst{1} \thanks{Corresponding author}
\and Xi Wu\inst{3} 
\and Xin Wang\inst{4} \thanks{Corresponding author} 
}

\institute{School of Computing and Artificial Intelligence, Southwest Jiaotong University \email{huangpeng@my.swjtu.edu.cn} \email{\{bpeng,jszhang\}@swjtu.edu.cn}
\and Department of Computer and Information Technology, Purdue University \email{hu968@purdue.edu}
\and School of Computer Science, Chengdu University of Information Technology  \email{wuxi@cuit.edu.cn}
\and University at Albany, State University of New York \email{xwang56@albany.edu}
}
\authorrunning{P. Huang et al.}

%
\maketitle              
\begin{abstract}
Current medical image classification efforts mainly aim for higher average performance, often neglecting the balance between different classes. This can lead to significant differences in recognition accuracy between classes and obvious recognition weaknesses. Without the support of massive data, deep learning faces challenges in fine-grained classification of fatty liver. In this paper, we propose an innovative deep learning framework that combines feature decoupling and adaptive adversarial training. Firstly, we employ two iteratively compressed decouplers to supervised decouple common features and specific features related to fatty liver in abdominal ultrasound images. Subsequently, the decoupled features are concatenated with the original image after transforming the color space and are fed into the classifier. During adversarial training, we adaptively adjust the perturbation and balance the adversarial strength by the accuracy of each class. The model will eliminate recognition weaknesses by correctly classifying adversarial samples, thus improving recognition robustness. Finally, the accuracy of our method improved by 4.16\%, achieving 82.95\%. As demonstrated by extensive experiments, our method is a generalized learning framework that can be directly used to eliminate the recognition weaknesses of any classifier while improving its average performance. Code is available at \url{https://github.com/HP-ML/MICCAI2024}.

\end{abstract}
\section{Introduction}
\begin{figure} [ht]
    \centering
    \includegraphics[width=0.95\textwidth]{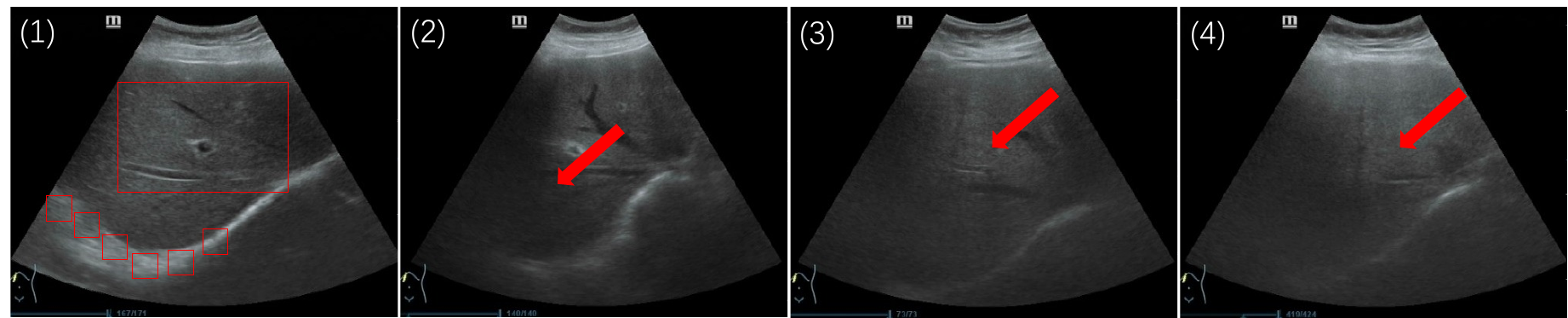}
    \caption{From left to right, the four ultrasound images are:(1) normal liver, (2) mild, (3) moderate, and (4) severe fatty liver. Squares mark the approximate location of the diaphragm and blood vessels. Arrows point to suspected locations of fat accumulation.}
    \label{fig:fattyliver}
\end{figure}

Deep learning has emerged as a pivotal tool, particularly in enhancing disease detection~\cite{review1}. However, deep learning for medical image classification faces challenges due to the scarcity of data for certain rare diseases~\cite{raredisease,lin2024robust,hu2023attention}. Additionally, much of the existing research focuses on achieving average optimal performance, with less emphasis on achieving balanced optimization across different classes~\cite{fattyliver,dlmui,hu2020learning,hu2022sum,hu2023rank}. Unlike natural image classification, where the objectives are clear and distinctions between classes, are evident. Different disease conditions are often classified based on subtle features \cite{wang2024artificial}. This makes the extraction of diverse features for classification more complicated. Fig.~\ref{fig:fattyliver} shows an example of a fatty liver ultrasound image. Hence, we interpret unbalanced samples to encompass not just data disparity but also the varying difficulty in recognition. 

Fatty liver is a disease that is usually asymptomatic and reversible in its early stages~\cite{fattyliver1}. Without timely and accurate diagnosis, the deterioration of fatty liver may lead to exacerbated liver damage, and even liver cancer. Ultrasonography is favored in fatty liver detection because of its cost-effectiveness and widespread availability~\cite{fattyliver2}. Distinguishing between different conditions of fatty liver requires careful examination of relevant features~\cite{ZHANG2023104073}. However, ultrasound images are often unable to directly distinguish between these stages and can only be described with vague adjectives such as "more", which complicates accurate identification. In order to overcome the problem of abundant but low quality information that makes it difficult for deep learning models to extract key pathology features, scholars have proposed many feature decoupling methods~\cite{Mu_Tang_Tan_Yu_Zhuang_2022,wang2020,yan2023}. These methods can guide us on how to make efficient use of fatty liver ultrasound images. 

Additionally, the difficulty of recognizing fatty liver disease varies from one period to another. When the model learning process is supervised only by the average performance, it may focus on identifying only certain fatty liver conditions and ignore hard-to-identify samples~\cite{ZHANG2021106448}. This learning process may increase the likelihood of an incorrect fatty liver diagnosis, which may hinder the medical diagnostic process.  In dealing with unbalanced performance, recent research has focused on the challenges posed by the long-tailed distribution of data ~\cite{FoPro-KD,WU2024108106,review2}, with less attention paid to the model's learning to a more intrinsic and robust feature representation.
Adversarial training can improve the model's robustness and enable the model to learn this feature representation~\cite{madry2017towards,Zhu2022,hu2024outlier}. In parallel, the model needs to generate smoother decision boundaries to correctly categorize adversarial samples~\cite{Wei2023,balaji2019instance,hu2021tkml,xie2023attacking}.
To address the problem of low-quality images with complex features in fatty liver ultrasound images, as well as the imbalance in model performance between classes. We propose a feature decoupling combined with adaptive adversarial training for fatty liver disease detection. The major contributions are as follows:
\begin{itemize}
    \item [1)]We propose a novel deep learning model, \textbf{I}terative \textbf{C}ompression \textbf{F}eature \textbf{D}ecoupler(ICFDNet), designed for feature decoupling. ICFDNet boosts sample utilization and ensures comprehensive information extraction and use.
    \item [2)]We adopt a color space transformation and selection strategy, which enriches the dataset and minimizes noise interference in model optimization.
    \item [3)]We utilize adversarial training to find the model's weaknesses and use it to generate more robust decision boundaries. Furthermore, the perturbation and adversarial strength are adaptively adjusted by the accuracy rate of each class, achieving an overall improvement in the model's capabilities.
\end{itemize}
\begin{figure} [t]
    \centering
    \includegraphics[width=0.95\textwidth]{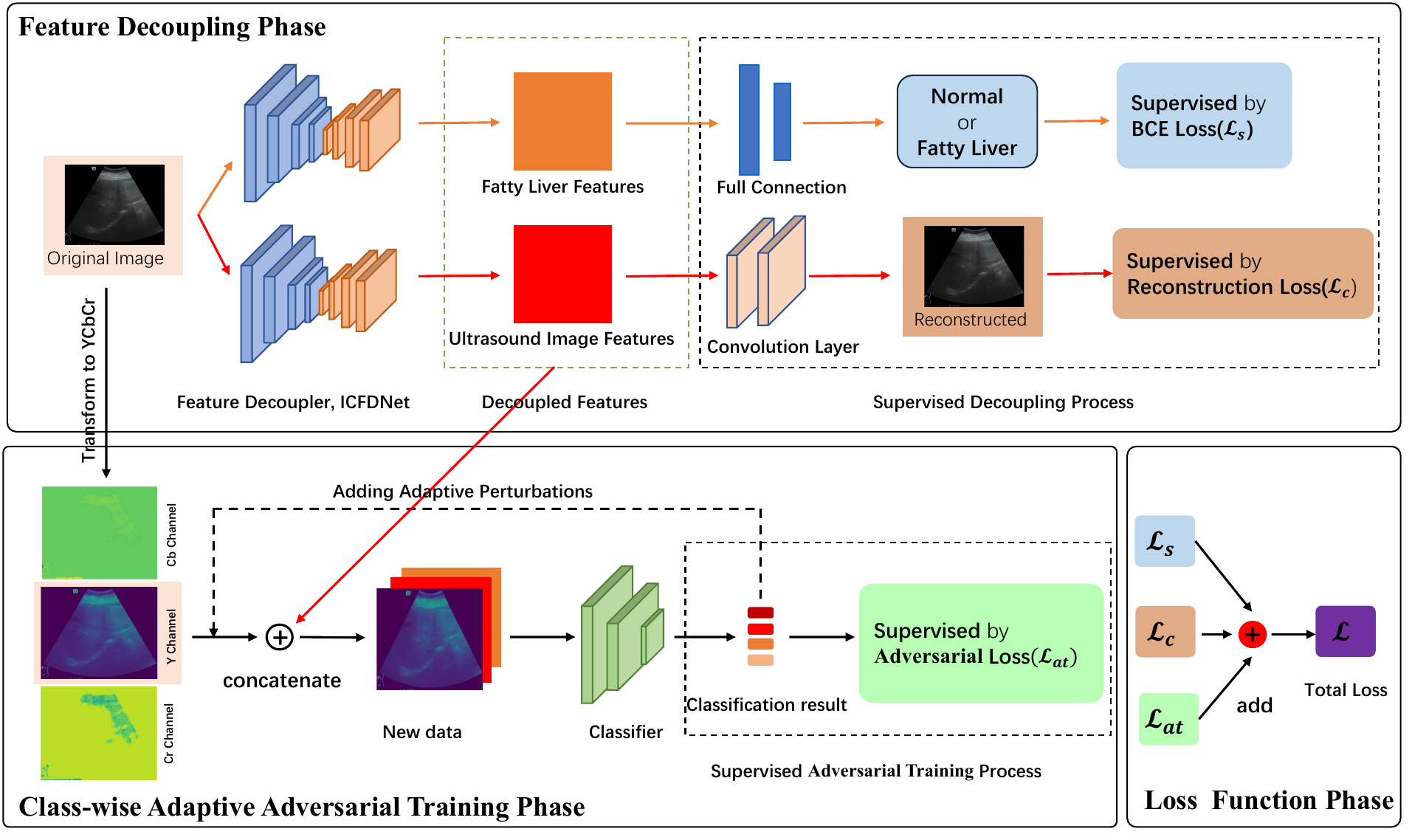}
    \caption{Overview of the proposed framework. (1) For the feature decoupling phase, we utilize the proposed ICFDNet to generate decoupled features and supervise the decoupling process in different methods. (2) For the adversarial training phase, we use an adaptive adversarial process to add corresponding perturbations to different images, and concatenate them with the decoupled features.}
    \label{fig:freamwork}
\end{figure}

\section{Proposed Methodology}
This section will describe the entire workflow and implementation details. Fig.~\ref{fig:freamwork} shows an overview of our proposed method. 
\subsection{Iterative Compression Feature Decoupler, ICFDNet}
\begin{figure} [t]
    \centering
    \includegraphics[width=0.95\textwidth]{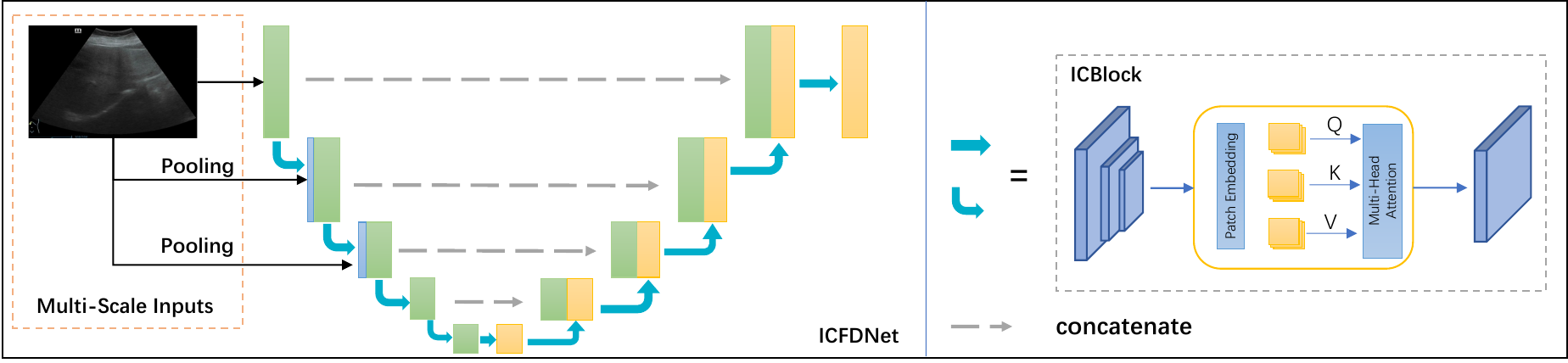}
    \caption{The network structure of ICFDNet. The network structure of ICFDNet utilizes a \textit{U} shaped architecture, with each of its ICBlocks, adopting a mini \textit{U} shaped structure. This design ensures that the network undergoes a complete \textit{compression-recovery} process during each operation, aiming to refine key features repeatedly.}
    \label{fig:ICFDNet}
\end{figure}

Fatty liver ultrasound images have high information density and insignificant feature differences. Therefore, the ICFDNet decoupler's goal is to decouple complex features into simple ones after learning. The structure of ICFDNet is shown in the Fig.~\ref{fig:ICFDNet}. The ICFDNet architecture is consistent with U-Net~\cite{U-net,tsai2024uu}. This \textit{U} pattern facilitates the extraction and decoupling of feature information from images. At the same time, we draw inspiration from the piston movement and realize that valuable information may not be effectively extracted by one compression process. Thus, a mini \textit{U} structure is used for all ICBlocks in ICFDNet. More detailed features are extracted by the model after iterative compression. Considering the importance of detail features in medical image recognition, this iterative compression pattern is particularly suitable for feature decoupling in fatty liver ultrasound image analysis.
Within the ICBlock module, the learning process consists of two phases: encoding and decoding. Initially, the module gradually downsamples the ultrasound image and captures local features through multiple convolution layers. Subsequently, the generated feature maps are passed through a multi-head self-attention module. The attention weights of important features are dynamically adjusted in the global view. Finally, these detailed features are enhanced by dilation convolution. Additionally, to prevent information loss due to the rapid compression caused by the relatively shallow depth of the ICBlock, ICFDNet employs multi-level pooling to provide multi-scale information. It combines initial information with deep features to ensure that key features are not forgotten throughout the process.
\subsection{Accuracy-based Adaptive Adversarial Training}
To improve the average performance of the classifiers and balance the inter-class performance, we utilize adaptive adversarial training to eliminate the recognition weaknesses of the classifiers. Adversarial samples are generated by adding perturbations to the original image that are not visible to the naked eye~\cite{at1}. These perturbations will help the model to generate more robust decision boundaries in the process of making mistakes. 

During adversarial training, ultrasound image samples of fatty liver with different conditions have different sensitivities to perturbations. And, it has been demonstrated that different classes should be added with different perturbations to achieve optimal results~\cite{Wei2023}. So, in order for the model to have the ability to learn more essential features. We utilize the accuracy of different categories to adaptively add perturbations to different categories. Its formula is expressed as Eq.~\ref{eq:adpertu}:
\begin{equation}\label{eq:adpertu}
    \epsilon_i = (\sigma + Acc_i ) \cdot \epsilon ,
\end{equation}
where $\epsilon$ represents the initial perturbation, and $Acc_i$ denotes the accuracy of class $i$ in the training set following the latest training epoch. The objective for the subsequent epoch is to adjust the perturbation $\epsilon_i$ for class $i$, based on $Acc_i$. To prevent $\epsilon_i$ from becoming too small and impeding the optimization process, we introduce a hyperparameter $\sigma$ to regulate the minimum perturbation level.

\begin{algorithm}[t]
\caption{Feature Decoupling and Adaptive Adversarial Training Optimization Process.}
\label{alg:1}
\begin{algorithmic}
\State{\textbf{Inputs:}} Model $M(\cdot)$ (including two ICFDNet: $IC_1{(\cdot)}$, $IC_2{(\cdot)}$; a classifier $f(\cdot)$); Train dataset $D=(x_i, y_i)$; Perturbation $\epsilon$, regularization parameter $\beta$; Weights $\lambda_1,\lambda_2,\lambda_3$.
\State{\textbf{Outputs:}} A performance-balanced and robust model $M(\cdot)$.
\State{Initialize} {$\epsilon_i \leftarrow \epsilon, \beta_i \leftarrow \beta$}
\For{$epoch \in Epochs$}
\For{$minibatch$}
\State $Feat_1, ~\hat{pred_i} \leftarrow IC_1(x_i)$, $Feat_2, ~\hat{x} \leftarrow IC_2(x)$; \Comment{Feature decoupling}
\State $Y_c \leftarrow x$, $x_c \leftarrow cat(Feat_1, Feat_2, Y_c)$; \Comment{Transformation and concatenation}
\State $x^\prime \leftarrow \max_{\|x^{\prime}-x\|\leq\epsilon_i} \mathcal{KL}(f(x_c), f(x^{\prime}))$; \Comment{Adversarial sample generation}
\State $l_c \leftarrow \mathcal{L}_c(\hat{x},x), l_s \leftarrow \mathcal{L}_s (pred_i,\hat{y_i}), l_{at} \leftarrow \mathcal{L}_{at}(f(x_c),f(x^\prime))$;\Comment{Loss}
\State $l \leftarrow \lambda_1l_c+\lambda_2 l_s+\lambda_3 l_{at}$;
\EndFor
\State $Acc_i \leftarrow f(x^\prime)$;\Comment{Update parameters}
\State $\epsilon_i \leftarrow Acc_i$, $\beta_i \leftarrow Acc_i$;
\EndFor
\State \Return $M(\cdot)$.
\end{algorithmic}
\end{algorithm}

With the appropriate $\epsilon_i$, we enable the model to capture more generalized fatty liver characteristics across classes. The likelihood of inter-class confounding is reduced. With our approach, the model not only obtains the basic features of each class using the original samples but also identifies the more essential features through the adversarial samples. However. the model may pay too much attention to the confrontation samples and ignore the original samples during the confrontation training process~\cite{tradeoff1,tradeoff2}. The converse also holds true. This situation may lead to a decrease in model performance.
Therefore, we refer to TRADES~\cite{zhang2019theoretically} to add the regularization term parameter $\beta_i$ to balance it. The calculation formula for $\beta_i$ is written as Eq.~\ref{eq:classbeta}:
\begin{equation}
    \label{eq:classbeta}
    \beta_i = \frac{(\mu + Acc_i) \cdot \beta}{1+(\mu + Acc_i) \cdot \beta} ,
\end{equation}
The meaning of $\mu$ is consistent with $\sigma$, and it is also a hyperparameter. It can be found that we compute $\beta_i$ for different results by $Acc_i$ to achieve adaptive customized weights between classes. The calibrated loss function $\mathcal{L}_{at}$ of the classifier module can be expressed as Eq.~\ref{eq:accloss}:
\begin{equation}
    \label{eq:accloss}
    \mathcal{L}_{at} = \underbrace{(1-\beta_i) \cdot \mathcal{L}(f(x),Y)}_{\mathcal{L}_{clean}} + \underbrace{\beta_i \cdot \max_{\|x^{\prime}-x\|\leq\epsilon_i}\mathcal{KL}(f(x), f(x^{\prime}))}_{\mathcal{L}_{robust}} ,
\end{equation}
where $f(x)$ is the output vector of classifier, $Y$ is the label vector, $\mathcal{L}(\cdot)$ is Cross-entropy loss, and $\mathcal{KL(\cdot)}$ is the KL divergence. During the training process, when the accuracy of some classes is lower, the corresponding $\beta_i$ is lowered. And the natural loss $\mathcal{L}_{clean}$ of these classes then receives a higher weight. This approach allows the model to pay more attention to the original features of natural samples, thus improving initial performance. With the increase in accuracy, the model gradually begins to focus on hard samples that are prone to confusion and clarifies the features of these samples through perturbations.
\subsection{Loss Function}
Our learning process consists of two parts: decoupled feature generation and classifier adversarial optimization. Alg.~\ref{alg:1} shows the simplified algorithm. It can be noticed that we have utilized three loss functions to jointly supervise the model learning process. The total loss $L$ is defined as Eq~\ref{eq:loss}:
\begin{equation}
    \label{eq:loss}
    \mathcal{L}=\lambda_1\mathcal{L}_c+\lambda_2 \mathcal{L}_s+\lambda_3 \mathcal{L}_{at} ,
\end{equation}
where $\lambda_1$, $\lambda_2$, $\lambda_3$ are the weight factors to control the impact of different parts on the whole task. $\mathcal{L}_c$ and $\mathcal{L}_s$ supervise the decoupling process. $\mathcal{L}_s$ is \textit{BCELosss}. The formula for $\mathcal{L}_c$ is written as Eq.~\ref{eq:commonloss}:
\begin{equation}
    \label{eq:commonloss}
    \mathcal{L}_c = \sqrt{(\Delta^2(x) - \Delta^2(\hat{x})) + \xi^2} + \mathcal{L}_1 ,
\end{equation}
Where $\Delta^2$ is the Laplace operator, $\hat{x}$ represents the reconstructed ultrasound image, and $\xi$ is a small constant used to ensure numerical stability. The important features of ultrasound images are mainly centered on ambiguous textures and boundaries. And Laplace operator is a second-order differential operator. It is capable of focusing on rapidly changing regions in ultrasound images~\cite{lvrnet}. Therefore, in the feature decoupling process of the fatty liver ultrasound image, using $\mathcal{L}_c$ can supervise the decoupler to obtain the boundary and texture information. The decoupler can better extract the high-frequency detail features. Meanwhile, the $\mathcal{L}_1$ loss function can better keep the brightness and color invariant and does not over-penalize the error. It further ensures better feature decoupling results.
\section{Experiment}



\begin{table} [t]
\caption{Accuracy performance of the baseline classifiers and ours in classifying ultrasound images of fatty liver. The \textit{\uuline{Best}} and \textit{\underline{Worst}} results for different classes are marked in italics. The \textbf{Best Average Performance} and \textbf{Smallest Gap} are highlighted.}
\label{tab:classifier_result}
\centering
\scalebox{0.9}{
\begin{tabular}{@{}ccccccc@{}}
\toprule
\multicolumn{1}{l|}{Method} & Mild\% & Moderate\% & Severe\% & \multicolumn{1}{c|}{Normal\%} & Best-Worst\% & Average\% \\ \midrule
\multicolumn{1}{l|}{ViT~\cite{ViT}}     & 53.33 & \textit{\underline{18.18}} & \textit{\uuline{92.30}} & \multicolumn{1}{c|}{86.96} & 74.12 & 59.09 \\
\multicolumn{1}{l|}{ResNet~\cite{ResNet}}  & 68.89 & \textit{\underline{63.64}} & 82.05 & \multicolumn{1}{c|}{\textit{\uuline{95.65}}} & 32.01 & 76.13 \\
\multicolumn{1}{l|}{CBAM~\cite{CBAM}}    & \textit{\underline{66.45}} & 75.76 & 84.62 & \multicolumn{1}{c|}{\textit{\uuline{97.10}}} & 30.65 & 78.79 \\
\multicolumn{1}{l|}{SENet~\cite{SENet}}   & \textit{\underline{63.33}} & 78.79 & 84.62 & \multicolumn{1}{c|}{\textit{\uuline{95.65}}} & 32.32 & 78.78 \\
\multicolumn{1}{l|}{SqueezeNet~\cite{SqueezeNet}} & 80.00 & \textit{\underline{57.58}} & 69.23 & \multicolumn{1}{c|}{\textit{\uuline{92.75}}} & 35.17 & 76.14 \\ 
 \midrule
\multicolumn{1}{l|}{Ours} & \textit{\underline{76.67}} & 77.27 & 84.62 & \multicolumn{1}{c|}{\textit{\uuline{95.65}}} & \textbf{18.98} & \textbf{82.95} \\ \bottomrule
\end{tabular}
}
\end{table}

\noindent\textbf{Dataset.} The data were sourced from the physical examination information of elderly fatty liver patients aged 65 and above who visited the Tiaodenghe Community in Chenghua District, Chengdu City, from 2020 to 2022. To ensure data quality, patients with blurry images due to massive liver occupancies, aerogenic interference, or obese physique were excluded, ultimately selecting data from 1265 patients. The ultrasound image diagnostic results were reviewed and confirmed by two senior physicians.

\noindent\textbf{Comparing with the Existing Methods.} Many deep learning studies on medical imaging utilize the best performing network as the underlying backbone. Special problems are targeted for modifications to achieve superior results. Consequently, we selected a deep learning classification model with strong generalization capabilities as the base model. This selection strategy might not prioritize the most recent network models, but it favors those with enhanced generalization performance.

Table~\ref{tab:classifier_result} shows the accuracy performance of different classifiers for the fatty liver recognition task. The results show that the baseline's performance varies significantly between classes, and there are obvious shortcomings in the recognition. It is worth celebrating that our method not only outperforms the other methods in terms of average performance, achieving 82.95\% but also obtains a balanced performance in fatty liver classification. The \textit{best-worst} performance difference was significantly reduced, achieving 18.98\%. Meanwhile, the recognition performance of our method for hard-to-recognize classes (e.g., mild and moderate fatty liver) was significantly improved. The obvious recognition weaknesses present in the baseline model were also effectively addressed.
From the resulting values, baseline's \textit{best-worst} performance spreads are all over 30\%, with the lowest being 30.65\% for the CBAM model. Additionally, the average accuracy of these baselines is not outstanding, with none of them exceeding 80\%. In particular, ViT has the worst performance among all the methods. Not only does it have the lowest average recognition accuracy, but also the performance difference between the best and worst performing classes is 74.12\%. \textit{Ours} improves the average performance by 4.16\% compared to the second-best CBAM, and the \textit{best-worst} performance difference decreases by 11.67\%. The experimental results demonstrate that the Ours approach may be more reliable in practical applications. This is due to its ability to provide consistent prediction across a wide range of cases without overall performance fluctuations due to specific classes.

\begin{table}[ht]
\caption{Ablation experiment results. \textbf{Best} in each group are highlighted; $\uparrow$ and $\downarrow$ represent accuracy improvements and decreases respectively; AT: \textbf{A}dversarial \textbf{T}raining.}
\label{tab:clean_result}
\centering
\scalebox{0.9}{
\begin{tabular}{@{}lccccl@{}}
\toprule
\multicolumn{1}{l|}{Method} & Mild\% & Moderate\% & Severe\% & \multicolumn{1}{c|}{Normal\%} & Average\%  \\ \midrule
\multicolumn{1}{l|}{ResNet~\cite{ResNet}}  & 68.89 & 63.64 & 82.05 & \multicolumn{1}{c|}{95.65} & 76.13 \\
\multicolumn{1}{l|}{ResNet+ICFDNet}  & 73.33 & 83.33 & 84.62 & \multicolumn{1}{c|}{84.06} & 80.30 $\uparrow$ \\
\multicolumn{1}{l|}{ResNet+ICFDNet+AT}   & 72.22 & 84.85 & 79.49 & \multicolumn{1}{c|}{88.41} & \textbf{80.68} $\uparrow$ \\ \midrule

\multicolumn{1}{l|}{CBAM~\cite{CBAM}}    & 66.45 & 75.76 & 84.62 & \multicolumn{1}{c|}{97.10} & 78.79 \\
\multicolumn{1}{l|}{CBAM+ICFDNet}    & 72.22 & 78.79 & 87.18 & \multicolumn{1}{c|}{86.96} & 79.93 $\uparrow$\\
\multicolumn{1}{l|}{CBAM+ICFDNet+AT}    & 73.33 & 81.82 & 76.92 & \multicolumn{1}{c|}{92.78} & \textbf{81.06} $\uparrow$\\ \midrule

\multicolumn{1}{l|}{SENet~\cite{SENet}}   & 63.33 & 78.79 & 84.62 & \multicolumn{1}{c|}{95.65} & 78.78 \\
\multicolumn{1}{l|}{SENet+ICFDNet}   & 75.89 & 82.33 & 84.53 & \multicolumn{1}{c|}{73.91} & 78.41 $\downarrow$\\ 
\multicolumn{1}{l|}{SENet+ICFDNet+AT}    & 74.44 & 84.85 & 84.62 & \multicolumn{1}{c|}{76.81} & \textbf{79.17} $\uparrow$\\ \midrule

\multicolumn{1}{l|}{SqueezeNet~\cite{SqueezeNet}} & 80.00 & 57.58 & 69.23 & \multicolumn{1}{c|}{92.75} & 76.14 \\ 
\multicolumn{1}{l|}{SqueezeNet+ICFDNet}  & 73.33 & 77.27 & 76.92 & \multicolumn{1}{c|}{91.30} & 79.55 $\uparrow$\\ 
\multicolumn{1}{l|}{SqueezeNet+ICFDNet+AT} & 76.67 & 77.27 & 84.62 & \multicolumn{1}{c|}{95.65} & \textbf{82.95} $\uparrow$\\ 
\bottomrule
\end{tabular}
}
\end{table}

\noindent\textbf{Ablation Study.} 
To demonstrate that our method can steadily improve the average performance of the classifier while still taking into account the performance balance between classes. We introduced different classification networks into our method and the experimental results are shown in Table~\ref{tab:clean_result}. From the experimental results, it can be observed that the performance of different classification networks has improved after employing the learning process we proposed. They all obtained the best results within the group. After the combination of SqueezeNet with ICFDNet, the average accuracy improved by 3.41\%. And after the addition of AT, it further increased by 3.4\% reaching 82.95\%. In addition, all of the significant recognition shortcomings present in the base classifiers were also improved considerably. In the moderate fatty liver, which is more difficult to recognize, ResNet improved its recognition accuracy by 21.21\%, SqueezeNet by 19.69\%, and CBAM and SENet by more than 6\%. We also discovered that the accuracy decreased by 0.37\% when SENet was only combined with ICFDNet, but when integrated with the entire training process, the accuracy was enhanced. This indicates that our approach has a certain level of complementarity. Taking into account all the experimental results, the method we proposed can effectively enhance all classification networks and eliminate their recognition weaknesses.

\section{Conclusion}
This paper introduces the distinctiveness of medical images compared to natural images in classification tasks. Then, we discuss existing work that prioritizes higher average performance while ignoring the performance balance between different classes. This leads to models with worse recognition performance for certain classes. In response, we decouple complex features using ICFDNet. And, we use adversarial training to eliminate the model's recognition weaknesses. Ultimately, our method not only improves the average classifier performance but also takes into account the performance balance between different classes. Additionally, our method can be directly adapted to any classifier, suitable for any situation. However, experiments have shown that while class balancing has improved, the accuracy of certain classes has decreased. In future work, we aim to find a method to achieve it without sacrificing the accuracy of other classes.
\begin{credits}
\subsubsection{\ackname} Supported by Natural Science Foundation of Sichuan 2022NSFSC0531, and Sichuan Science and Technology Program  2023YFG0125, 2024ZDZX0001, 2022YFG0029.

\subsubsection{\discintname}
The authors declare that the research was conducted in the absence of any commercial or financial relationships that could be construed as a potential conflict of interest.
\end{credits}

\bibliography{Paper-3325} 

\begin{thebibliography}{10}
\providecommand{\url}[1]{\texttt{#1}}
\providecommand{\urlprefix}{URL }
\providecommand{\doi}[1]{https://doi.org/#1}

\bibitem{Zhu2022}
Allen-Zhu, Z., Li, Y.: Feature purification: How adversarial training performs robust deep learning. In: FOCS. pp. 977--988 (2022). \doi{10.1109/FOCS52979.2021.00098}

\bibitem{balaji2019instance}
Balaji, Y., et~al.: Instance adaptive adversarial training: Improved accuracy tradeoffs in neural nets. arXiv preprint arXiv:1910.08051  (2019)

\bibitem{fattyliver1}
Brunt, E.M., et~al.: Nonalcoholic fatty liver disease. Nature reviews Disease primers  \textbf{1}(1),  1--22 (2015)

\bibitem{ViT}
Dosovitskiy, A., et~al.: An image is worth 16x16 words: Transformers for image recognition at scale. arXiv preprint arXiv:2010.11929  (2020)

\bibitem{FoPro-KD}
Elbatel, M., et~al.: Fopro-kd: Fourier prompted effective knowledge distillation for long-tailed medical image recognition. IEEE TMI  (2023)

\bibitem{fattyliver2}
Han, A., et~al.: Noninvasive diagnosis of nonalcoholic fatty liver disease and quantification of liver fat with radiofrequency ultrasound data using one-dimensional convolutional neural networks. Radiology  \textbf{295}(2),  342--350 (2020)

\bibitem{ResNet}
He, K., et~al.: Deep residual learning for image recognition. In: CVPR. pp. 770--778 (2016). \doi{10.1109/CVPR.2016.90}

\bibitem{SENet}
Hu, J., et~al.: Squeeze-and-excitation networks. In: CVPR. pp. 7132--7141 (2018). \doi{10.1109/CVPR.2018.00745}

\bibitem{hu2023attention}
Hu, J., et~al.: Attention guided policy optimization for 3d medical image registration. IEEE Access  (2023)

\bibitem{hu2020learning}
Hu, S., et~al.: Learning by minimizing the sum of ranked range. NeurIPS  (2020)

\bibitem{hu2021tkml}
Hu, S., et~al.: Tkml-ap: Adversarial attacks to top-k multi-label learning. In: ICCV. pp. 7649--7657 (2021)

\bibitem{hu2022sum}
Hu, S., et~al.: Sum of ranked range loss for supervised learning. JMLR  (2022)

\bibitem{hu2023rank}
Hu, S., et~al.: Rank-based decomposable losses in machine learning: A survey. IEEE TPAMI  (2023)

\bibitem{hu2024outlier}
Hu, S., et~al.: Outlier robust adversarial training. In: ACML (2024)

\bibitem{SqueezeNet}
Iandola, F.N., et~al.: Squeezenet: Alexnet-level accuracy with 50x fewer parameters and $<$0.5mb model size. arXiv:1602.07360  (2016)

\bibitem{raredisease}
Li, X., et~al.: Difficulty-aware meta-learning for rare disease diagnosis. In: MICCAI. pp. 357--366 (2020). \doi{10.1007/978-3-030-59710-8\_35}

\bibitem{lin2024robust}
Lin, L., et~al.: Robust covid-19 detection in ct images with clip. MIPR  (2024)

\bibitem{at1}
Liu, J., et~al.: Adversarial machine learning: A multilayer review of the state-of-the-art and challenges for wireless and mobile systems. IEEE COMST  \textbf{24}(1),  123--159 (2022). \doi{10.1109/COMST.2021.3136132}

\bibitem{madry2017towards}
Madry, A., et~al.: Towards deep learning models resistant to adversarial attacks. arXiv preprint arXiv:1706.06083  (2017)

\bibitem{Mu_Tang_Tan_Yu_Zhuang_2022}
Mu, Z., et~al.: Disentangled motif-aware graph learning for phrase grounding. In: AAAI. p. 13587–13594 (2022). \doi{10.1609/aaai.v35i15.17602}

\bibitem{lvrnet}
Pahwa, E., et~al.: Lvrnet: Lightweight image restoration for aerial images under low visibility. arXiv preprint arXiv:2301.05434  (2023)

\bibitem{ZHANG2023104073}
Pengfei, Z., et~al.: Feature analysis and automatic classification of b-mode ultrasound images of fatty liver. BSPC  \textbf{79},  104073 (2023). \doi{10.1016/j.bspc.2022.104073}

\bibitem{fattyliver}
Reddy, D.S., et~al.: A novel computer-aided diagnosis framework using deep learning for classification of fatty liver disease in ultrasound imaging. In: IEEE Healthcom. pp.~1--5 (2018). \doi{10.1109/HealthCom.2018.8531118}

\bibitem{U-net}
Ronneberger, O., et~al.: U-net: Convolutional networks for biomedical image segmentation. In: MICCAI. pp. 234--241. Springer (2015)

\bibitem{tradeoff1}
Tian, Q., et~al.: Analysis and applications of class-wise robustness in adversarial training. In: KDD. pp. 1561--1570 (2021)

\bibitem{tsai2024uu}
Tsai, T.Y., et~al.: Uu-mamba: Uncertainty-aware u-mamba for cardiac image segmentation. arXiv preprint arXiv:2405.17496  (2024)

\bibitem{tradeoff2}
Wang, H., Wang, Y.: Generalist: Decoupling natural and robust generalization. In: CVPR. pp. 20554--20563 (2023)

\bibitem{wang2020}
Wang, X., et~al.: Disentangled graph collaborative filtering. In: SIGIR. p. 1001–1010 (2020). \doi{10.1145/3397271.3401137}

\bibitem{wang2024artificial}
Wang, X., Zhu, H.: Artificial intelligence in image-based cardiovascular disease analysis: A comprehensive survey and future outlook. arXiv preprint arXiv:2402.03394  (2024)

\bibitem{dlmui}
Wang, Y., et~al.: Deep learning in medical ultrasound image analysis: A review. IEEE Access  \textbf{9},  54310--54324 (2021). \doi{10.1109/ACCESS.2021.3071301}

\bibitem{Wei2023}
Wei, Z., et~al.: Cfa: Class-wise calibrated fair adversarial training. In: CVPR. pp. 8193--8201 (2023). \doi{10.1109/CVPR52729.2023.00792}

\bibitem{CBAM}
Woo, S.e.a.: Cbam: Convolutional block attention module. In: ECCV. pp. 3--19 (2018)

\bibitem{WU2024108106}
Wu, Z., et~al.: Medical long-tailed learning for imbalanced data: Bibliometric analysis. CMPB p. 108106 (2024). \doi{10.1016/j.cmpb.2024.108106}

\bibitem{xie2023attacking}
Xie, Y., et~al.: Attacking important pixels for anchor-free detectors. arXiv preprint arXiv:2301.11457  (2023)

\bibitem{yan2023}
Yan, Z., et~al.: Ucf: Uncovering common features for generalizable deepfake detection. In: ICCV. pp. 22355--22366 (2023). \doi{10.1109/ICCV51070.2023.02048}

\bibitem{zhang2019theoretically}
Zhang, H., et~al.: Theoretically principled trade-off between robustness and accuracy. In: ICML. pp. 7472--7482. PMLR (2019)

\bibitem{ZHANG2021106448}
Zhang, R., et~al.: Mbnm: Multi-branch network based on memory features for long-tailed medical image recognition. CMPB  \textbf{212},  106448 (2021). \doi{10.1016/j.cmpb.2021.106448}

\bibitem{review2}
Zhang, Y., et~al.: Deep long-tailed learning: A survey. IEEE TPAMI  \textbf{45}(9),  10795--10816 (2023). \doi{10.1109/TPAMI.2023.3268118}

\bibitem{review1}
Zhou, S.K., et~al.: A review of deep learning in medical imaging: Imaging traits, technology trends, case studies with progress highlights, and future promises. Proceedings of the IEEE  (2021). \doi{10.1109/JPROC.2021.3054390}

\end{thebibliography}
\bibliographystyle{splncs04}
\end{document}